\documentclass[pra,twocolumn,aps]{revtex4}
\usepackage{graphics}

\begin{document}

\title{Information preserved guided scan pixel difference coding for medical images}

\author{Kunio Takaya}
\affiliation{Electrical Engineering, University of Saskatchewan, Saskatoon, CANADA}

\author{C. Tannous}
\altaffiliation{Present address: Laboratoire de Magnétisme de Bretagne, UPRES A CNRS 6135,
Université de Bretagne Occidentale, BP: 809 Brest CEDEX, 29285 FRANCE}
\author{Li Yuan}
\affiliation{Electrical Engineering, University of Saskatchewan, and TRLabs, Saskatoon, CANADA}

\date{March 28, 2001}


\begin{abstract}
This paper analyzes the information 
content of medical images, with 3-D MRI images 
as an example, in terms of information entropy. 
The results of the analysis justify the use of Pixel 
Difference Coding for preserving all information 
contained in the original pictures, lossless coding 
in other words. The experimental results also indicate that the compression ratio CR=2:1 can be 
achieved under the lossless constraints. A pratical implementation of Pixel Difference Coding 
which allows interactive retrieval of local ROI (Region of Interest), while maintaining the near low 
bound information entropy, is discussed.
\\


\end{abstract}

\maketitle

\section{Introduction}
This research attempts at developing an information preserving image coding suitable for handling medical images which usually do not allow any degree of image degradation in comparison with the source images \cite{saghri, takaya}. JPEG, one of the well accepted compression standards, exhibits an excellent compression gain, but does not insure that the decoded result is the exact replica of the original. A variation of DPCM or pixel difference coding, namely "Guided Scan Pixel Difference Coding  was studied. A major modification to the classical DPCM is the inclusion of scan information, in terms of the direction used to calculate the pixel difference and step size. This additional information does not significantly increase the volume of coded data. It actually opens the extensibility for the DPCM to be able to handle multi-dimensional image data, in a similar manner as JPEG is extended to MPEG to handle movie presentation. Another important asp ect of medical image coding, which is the ability to access the full details of a local image (often referred to as ROI, Region of Interest), is also satisfied. Neither the location of a ROI nor the scanning method is restricted so that a local area of ROI can be selected anywhere in the original image. The flexibility in scanning makes it possible to use the same code for real time movie presentation in the same way MPEG uses JPEG. This research considers two stage redundancy removal, one by taking pixel differences and the other by an entropic source coding similar to the Huffman coding to achieve a redundancy removal of approximately 4-5 bpp (bits per pixel).\\

The sizes of medical images have increased with the advancement in various medical imaging modalities, typically in MRI and X-ray CAT scan. MRI can scan a whole body within a reasonable time of 30 minutes and X-ray CAT has adopted helical scanning for a higher spatial resolution. In the case of 3D scanning, the image size could be as large as 100 million pixels. Images of this size are usually stored on writable laser disks (CD-WORM), but this imposes serious problems when the retrieval of such data is attempted through a communication channel in the wide area network (WAN) environment using 64 kbits/sec bit-rate telephone channels.\\

The problems associated with the compression of medical images are two fold. The first requirement is that compression must be absolutely non-degradable. Regardless of whether a picture has been affected by measurement noise that occurs in the process of physical measurement and in image reconstruction, the source data acquired at an imager must be retained without any kind of losses. The second condition is the integrity of the objects contained in a picture data. Physicians examine very closely the image in a ROI (Region of Interest) or a number of ROI's. A selected ROI needs to be immediately accessible and it must be translated into a serial stream of compressed data for transmission via a communication channel.

\section{Technical Background}
The method is best represented by the name  Guided Scan Pixel Difference Coding \cite{takaya}, a variation of DPCM widely used in compressing voice signals. It is well known that a PCM voice signal of 6-8 bits can be compressed down to a 3 bit DPCM. Even discounting the fact that DPCM uses a nonlinear quantization scale having a greater step size for a larger signal level, it is easy to remove 3 bits per code word, reducing the total volume down to half of the original. In comparison, JPEG performs a compression rate of 10 and is better than DPCM, because it permits the encoded image to degrade somewhat from the original. DPCM is a non-degradable source coding if the quantization step size is fixed to the same linear scale as already done in the source images. Furthermore, the encoding/decoding method is basically the simple operation of subtraction/summation. Further compression to remove the redundancy leftover from DPCM is accomplished by an entropy coding such as Huffman code.\\

In order to satisfy the other aspect of medical image compression, accessibility to a ROI, additional information describing how an individual code representing a pixel difference is derived, must be attached at the expense of compression gain. In the case of 2D images, there are four (up, down, left and right) or eight (if diagonal differences are considered) possible directions by which a pixel difference can be calculated. If an ROI is specified by a person viewing an image specifying several guide marks surrounding the ROI, the length of each raster along a specified scan direction must be registered. Also, the step size used to calculate the pixel difference must be included, if coarsely sampled images need to be transmitted for viewing an overall image at a reduced image quality. These additional pieces of information could be 4 bits for directions and about 3 bits to indicate the step size per picture. Information regarding a length of scanning depends on how irregular a shape is set to be an ROI. Nevertheless, no significant increase in the total volume of coded words is expected. The term "Guided Scan is meant to include the information as to how an image or a ROI is scanned. The way of scanning an image should be left for the image coding system to determine, so that the total volume of information can be minimized for individual selected ROI's. Another interesting aspect of the Guided Scan DPCM is the structural similarity observed in the forward or backward difference interpolation algorithm. Since the tree of high order differences, necessary for the interpolation, can be calculated easily from the first order differences coded by the Guided Scan DPCM, there is potential to artificially increase the spatial resolution of the source image without distorting the image.

\section{Information Entropy}
The information entropy is defined by:

\begin{equation}
H=-\sum_{i=1}^{I}p_{i} \mbox{ log}_{2} p_{i}
\end{equation}

where $p_{i}$ is the probability associated with the occurrence of a value i when the total number of values used is $I$. If eight bits are used to code an image and the probability for any number between 0 and 255 to occur is the same, the information entropy is 8. If only one value, say 128, occurs all times, the entropy is zero. Adjacent pixels in a digitized image are highly correlated. If two adjacent pixels are considered, the probability for the first pixel $x$ to take a value i, that for the second pixel $y$ to take j, and the joint probability for the pixel $x$ to take a value i and the pixel $y$ to take j are given respectively as follows:

\begin{equation}
p_{i}=\frac{N_{i}}{N}, p_{j}=\frac{N_{j}}{N} \mbox{ and } p_{ij}= \frac{N_{ij}}{N} 
\end{equation}

The entropies based on the joint probability and the conditional probabilities are then obtained as follows:
\begin{eqnarray}
H (x,y ) &=& -\sum_{i,j} p_{ij} \mbox{ log}_{2} p_{ij}  \nonumber \\
H (x|y ) &=& -\sum_{i,j} p_{ij} \mbox{ log}_{2} \frac{p_{ij}}{p_{j}}  \nonumber \\
H (y|x ) &=& -\sum_{i,j} p_{ij} \mbox{ log}_{2} \frac{p_{ij}}{p_{i}}  
\end{eqnarray}

We also know that $H (x, y ) = H (x) + H (y |x) = H (y) + H (x|y )$. Since theoretically:

\begin{equation}
H (y |x)  - H (y ) \le 0 
\end{equation}

The entropy calculated for the second pixel $y$ knowing the occurrence of the first pixel $x$ is smaller than the entropy calculated from $y$ alone. The uncertainty coefficient of $y$

\begin{equation}
U (y |x) = \frac{H(y)-H(y|x)}{H(y)}
\end{equation}

is indicative of the dependency of $y$ on $x$. The symmetrical uncertainty:

\begin{equation}
U (x,y) = \frac{H(x)U (x|y) +H(y)U(y|x)}{H(x)+H(y)}
\end{equation}

yields 0 if $x$ and $y$ are completely independent and 1 if they are completely dependent.

\section{Entropy Calculated from Stretched Exponential PDF}

When the probability density function (PDF), obtained from a histogram of pixel intensities, is well approximated by the stretched exponential probability density function \cite{mallat91, mallat92,mallat89}:

\begin{equation}
p(x) = K e^{- {(\frac{|x|}{\alpha})}^\beta}  \hspace{1cm} \mbox{ with } K=\frac{\beta}{ 2\alpha\Gamma(\frac{1}{\beta}) }
\end{equation}   

the theoretical entropy can be calculated from:

\begin{equation}
H=2K \frac{\alpha}{\beta}\frac{(-lnK)\Gamma(\frac{1}{\beta})+\Gamma(1+\frac{1}{\beta})}{ln2}
\end{equation}

After evaluating the  absolute mean $\mu$ and standard deviation $\sigma$, the parameters $\alpha$
 and $\beta$ are extracted from the equation \cite{press}:

\begin{equation}
F(\beta) =\frac{\mu^2}{{\sigma^{2}+\mu^2}}
\end{equation}

where, the function $F$ is defined by:

\begin{equation}
F(x) =\frac{{\Gamma(\frac{2}{x}) }^2}{\Gamma(\frac{1}{x})\Gamma(\frac{3}{x})}
\end{equation}

and $\alpha$ is defined by:

\begin{equation}
\alpha=\frac{ {(\sigma^{2}+\mu^2)} \Gamma(\frac{1}{\beta})}{\Gamma(\frac{3}{\beta})}
\end{equation}

Fig. 1 depicts the variation of the entropy $H$ as a function of $\alpha$ in the interval [0.1-5] for a given $\beta$ picked in the interval [0.8-1.5].

\section{Preliminary Studies on Information Entropy}
A 3-D MRI brain section image which consists of 64 slices was analyzed with an attempt to find the lower bound of the information entropy for medical images \cite{takaya}. The lower bound is a definite measure that tells exactly how many bits can be removed per pixel as redundancy. One slice of the 3-D image is shown in Fig. 2. Two different approaches are taken to calculate the entropies for the image shown in Fig. 2.

\begin{enumerate}

\item Use Eq. 1 which does not consider the dependency between adjacent pixels for n-th order difference 
images (n = 0,...,10).

\item Use Eq. 3 which takes the dependency of a pixel on its immediate neighbours into account (equivalent 
to a first order Markov model).
\end{enumerate}

The first approach finds a crude information entropy, if it is applied to the original picture since the redundancy due to pixel-to-pixel correlation is still intact. In order to find a more accurate information entropy, it is necessary to remove the redundancy by some means that allows the recovery of the original image. A difference image is introduced for this purpose. When a 2-D image is denoted by $A$ and $z^{-1}$ represents the one-step shift operator towards the positive side of the horizontal axis (or vertical or diagonal axis), the first difference is:

\begin{equation}
D_{1}=A-z^{-1}A=(1-z^{-1})A
\end{equation}

The second difference is given by
\begin{equation}
D_{2}=(1-z^{-1})D_{1}={(1-z^{-1})}^{2}A
\end{equation}

Thus the n-th  difference is given by
\begin{equation}
D_{n}={(1-z^{-1})}^{n}A
\end{equation}

In order to make reconstruction possible, the first column of the first difference must retains the first column of the original if horizontal shift is used. The second difference must retain the first column of the original and the second column of the first difference in its second column. Additional DC restoration columns must be progressively added when a new difference image is created. Successive applications of this difference operator remove the pixel-to-pixel correlation and resulting images become gradually more random. Table 1 shows how the entropy associated with such a difference image varies when n is increased.

\begin{table}[htbp]
\begin{center}
\begin{tabular}{ |c| c| }
\hline
Difference image  & Entropy \\
\hline
n =0 & 5.615405 \\
n=1  & {\bf 4.861321} \\
n=2 & 5.448454 \\
n=3 & 6.240144 \\
n=4 & 7.103492 \\
n=5 & 7.997103 \\
n=6 & 8.908238 \\
n=7 & 9.813683 \\
n=8 & 10.725728 \\
n=9 & 11.610693 \\
n=10 & 12.468448 \\
\hline
\end{tabular}
\caption{Entropies (bits/pixel) of Difference Images} \label{tab1}
\end{center}
\end{table}

2-D auto-correlation functions for the difference images of n= 0, 1 and 2 are shown in Fig. 3. The top figure is the 2-D auto-correlation of the original image. Comparing the middle n =1 and the bottom n =2, it is observed that the central peak of n=2 is sharper (less correlated) than that of n=1. As for the entropies calculated, the entropy drastically drops to the minimum at the very first difference operation then it starts increasing as n increases. Successive difference operations seem to decorrelate the image and make it more random, but the entropy monotonously increases after the first difference. This phenomenon can be explained from the frequency response of the n-th difference operation described by the transfer function of a high pass filter,

\begin{equation}
G(z)={(1-z^{-1})}^{n}
\end{equation}

Since the magnitude response is given by:

\begin{equation}
|G(e^{j\omega})|=2^{n} sin^{n}( \frac{\omega}{2} ) \hspace{1cm} \mbox{ where } 0 \le \omega < \pi
\end{equation}

and the bandwidth becomes wider as n increases. The power of the n-th difference image is greater than that of (n-1)th difference image, and so is the variance.\\

The second approach using Eq. 3 considers the dependency between adjacent pixels in calculating information entropies so that it is no longer necessary to decorrelate images. Since the first difference achieves the minimum entropy value, Eq. 3 is applied to the image of the first difference. Table 2 summarizes the results. The entropy $H (x , y )$ based on the joint probability $p_{ij}$
is divided by 2 to translate it into the entropy per pixel. This value is significantly smaller than $H (x) \mbox{ or } H (y )$.\\

Another interesting result is found by fitting the stretched exponential PDF \cite{daub} to the histogram of the original image shown in Fig. 4. The entropy for the large peak located in the lower gray scale range of $< 50$ is calculated to be 2.749977 without including the range $> 50$. This is nearly one half of the total entropy found for the original image. A whole picture of a medical image usually contains a significant portion of dark background.

\begin{table}[htbp]
\begin{center}
\begin{tabular}{ |c| c| }
\hline
 Entropy & value \\
\hline
$H(x)$  & 4.861343 \\
$H(y)$ &  4.861343  \\
$H(x, y)$  & 9.159570 \\
$H(x, y)/2$ &  4.579785  \\
$H(y|x)$   & 4.298226 \\
$H(x|y)$  &  4.298226 \\
\hline
\end{tabular}
\caption{Information Entropies Considering Adjacent Pixel-to-pixel Correlation} \label{tab2}
\end{center}
\end{table}

\section{Discussion}
 Reviewing the experimental results presented in the previous sections, some conclusive remarks can be made to determine the strategy for information preserving source coding. Information preserving coding or lossless coding generally means that the picture received is the picture sent in terms of the bit structure of the image, not in terms of the visual impression of the image before and after image compression/transmission. In this narrow sense of lossless coding, altering pixels is prohibited. No alterations in quantization are 
permitted. With these strict constraints, the only possible source of redundancy that could be removed is limited to the pixel-to-pixel correlation. Pixel difference coding (DPCM) alone brings the information entropy down to its near minimum. As seen in the entropy $H(x, y )$, the result of the pixel difference coding can be further trimmed but not to a large extent. The MR images analyzed are all 8 bit images. According to the sample calculations shown in this paper and other tested results, a bare-bone information entropy per pixel is slightly greater than 4 bits/pixel. An optimistic compression ratio CR is therefore CR=2:1, as long as a near optimum compression algorithm, typically Huffman coding, is used. Blending the Huffman code and the codes which consider state transitions that frequently occur within a near zero range of the pixel difference scale, a small improvement in compression ratio will be made. For example, a run of successive zeros up to a length of 10 can be treated as a code word if the frequencies of such occurrences are sufficiently high.\\

Further improvement of CR requires removing a constraint on quantization. As used in DPCM in voice signal coding, there are several methods to set up a quantization scale which minimizes the information entropy. If the methods of transform coding are allowed, CR can be improved drastically. Approaches to control the quality of medical images, for example texture, appearance of speckles, maintaining repeated basic patterns as fractal images do, etc... If one coding method can assure the fidelity of a certain image attribute, it might be considered as a better coding scheme. The medical community may be prepared to accept it as a better replacement for the totally reconstructible information preserving coding.\\

If some loss is permitted in medical image coding, the n-th order difference images discussed in this paper has another potential application for image coding. Recalling that the power spectrum $S (\omega_{x}, \omega_{y})$ of an n-th order difference image $D_{n}(x,y)$ relates to its auto-correlation $R(x,y)$ with the 2-D Fourier transform,

\begin{eqnarray}
&& S (\omega_{x}, \omega_{y})  = \mathcal{F} R(x,y)   \nonumber \\
&& \hspace{2cm}                    = \{\mathcal{F} D_{n}(x,y) \} \{\mathcal{F} D_{n}(x,y) \}^{\ast}
\end{eqnarray}

it is apparent that the magnitude information of $D_{n}(x,y)$ is contained in the highly concentrated peak of $R(x,y)$. Since the phase information is lost in calculating $R(x,y)$, it is necessary to preserve the phase of $D_{n}(x,y)$. The auto-correlation of $\angle \mathcal{F} D_{n}(x,y)$  shown in Fig. 5 indicates that the magnitude of phase is also highly concentrated at zero of the 2-D phase auto-correlation. The n-th order difference image produces a highly concentrated auto-correlation both in magnitude and in phase. By preserving the profile of the peak and discarding the rest, it seems possible to achieve a high compression gain. It is however known that image degradation is usually enhanced by inaccurate phase estimation.\\

{\bf Acknowledgement} 
Li Yuan acknowledges support through a TRlabs fellowship. C. Tannous acknowledges support from NSERC president's fund grant.

\vspace{1cm}
\centerline{\Large\bf Figure Captions}
\vspace{1cm}

\begin{itemize}

\item[Fig.\ 1:] Theoretical entropy calculated for stretched exponential
 probability density functions. The graphs show the entropy versus $\alpha$ with
 $\beta$ as a parameter.

\item[Fig.\ 2:]  A Slice of a 3-D MRI Head Image (256x256)

\item[Fig.\ 3:] 2D Auto-correlation, original, 1st difference, 2nd difference image and phase image of 
the original (from top to bottom). Top vertical scale is in units of $10^7$ , middle is in units of $10^5$  whereas
bottom part is in $10^6$ units.

\item[Fig.\ 4:] Histograms for the original, 1st difference and 2nd difference images.

\item[Fig.\ 5:] Auto-correlation of the phase image. Vertical scale is in units of $10^4$.

\end{itemize}

\end{document}